\documentclass[aps,pre,twocolumn,showkeys]{revtex4}
\usepackage{graphicx}
\usepackage{amssymb,amsfonts,amsmath}

\def\jb{{\bf j}}

\def\ub{{\bf u}}
\newcommand{\Dukhin}[1][]{\mathrm{\hspace{1pt}D\hspace{-1pt}u{#1}}\hspace{1pt}}
\def\Du{{\tilde{\rho}_s}}

\begin{document}

%\preprint{}

\title{Desalination shocks in microstructures}

% \email, \thanks, \altaffiliation all apply to the current
% address or url should go in the {}'s for \email.

\author{Ali Mani}
\email[]{alimani@mit.edu}
\affiliation{Department of Chemical Engineering, Massachusetts Institute of Technology, Cambridge, MA 02139}
 \author{Martin Z. Bazant}
\email[]{bazant@mit.edu}
\affiliation{Department of Chemical Engineering, Massachusetts Institute of Technology, Cambridge, MA 02139}
\affiliation{Department of Mathematics, Massachusetts Institute of Technology, Cambridge, MA 02139}

%\thanks{}


%\date{\today}

\begin{abstract}
Salt transport in bulk electrolytes is limited by diffusion and convection, but in microstructures with charged surfaces (e.g. microfluidic devices, porous media, soils, or biological tissues) surface conduction and electro-osmotic flow also contribute to ionic fluxes. For small applied voltages, these effects lead to well known linear electrokinetic phenomena. In this paper, we predict some surprising nonlinear dynamics that can result from the competition between bulk and interfacial transport at higher voltages. When counter-ions are selectively removed by a membrane or electrode, a ``desalination shock" can propagate through the microstructure, leaving in its wake an ultrapure solution, nearly devoid of co-ions and colloidal impurities. We elucidate the basic physics of desalination shocks and develop a mathematical theory of their existence, structure, and stability, allowing for slow variations in surface charge or channel geometry.  Via asymptotic approximations and similarity solutions, we show that desalination shocks accelerate and sharpen in narrowing channels, while they decelerate and weaken, and sometimes disappear, in widening channels. These phenomena may find applications in   separations (desalination, decontamination, biological assays) and energy storage (batteries, supercapacitors) involving electrolytes in microstructures. 
\end{abstract}


\maketitle

\section{Introduction}
\label{intro}

All electrochemical processes lead to ionic concentration gradients in electrolytes ~\cite{probstein1994,newman2004}. In water desalination, the removal of ions is the desired outcome, but in most other situations, such as energy storage by batteries or energy conversion by fuel cells, salt depletion is undesirable because it increases the solution resistance and slows electrochemical reactions, thereby increasing the over-potential required to maintain a desired current. Salinity variations also commonly arise in biological systems due to the action of membranes or external stimuli, and their dynamics can significantly affect living cells and tissues. In all of these situations it is important to understand the dynamics of ions in complex geometries.

It is generally assumed that salt transport in bulk electrolytes occurs only by diffusion and convection. This hypothesis underlies important industrial processes, such as electrodialysis~\cite{sonin1968,nikonenko2010}, electrodeposition~\cite{rosso2007}, and experimental techniques, such as impedance spectroscopy~\cite{barsoukov2005}, cyclic voltammetry~\cite{bard1980}. In a concentrated electrolyte, ionic diffusion is nonlinear (with a concentration-dependent diffusivity~\cite{newman2004}), but the familiar {\it square-root of time} scaling of linear diffusion usually remains~\cite{crank1975}. This conclusion also holds for macroscopic transport in porous media, as long as linear diffusion occurs within the pores~\cite{Torquato:2002}.

Recent experiments have shown that more complicated, nonlinear dynamics are possible if strong salt depletion (``concentration polarization") occurs in microstructures. A growing body of work has focused on Dukhin's second-kind electro-osmotic flows~\cite{Dukhin:1991,Mishchuk:2001} and the Rubinstein-Zaltzman instability~\cite{Rubinstein:2000,Zaltzman:2007} near electrodialysis membranes~\cite{rubinstein2008,nikonenko2010} and microchannel/nanochannel junctions~\cite{kim2007,Yossifon:2008} and in packed beds of particles~\cite{Leinweber:2004,Tallarek:2005}. In all of these cases, the transport of ions across a selective surface depletes the salt concentration and causes nonlinear electrokinetic phenomena in electric double layers (EDLs) sustaining {\it normal} current.  

In contrast, our focus here is on the effect of {\it tangential} current in the EDL~\cite{Lyklema:1995,Chu:2007,khair2008}, also known as ``surface conduction", which has a long history, prior to microfluidics~\cite{smoluchowski1905,bikerman1935,urban1935,overbeek1943,deryagin1969}. 
In linear electrokinetics, the importance of surface conduction is controlled by the Dukhin number~\cite{Lyklema:1995,bikerman1940},
\begin{equation}
\Dukhin = \frac{\kappa'_s }{\kappa_b h},
\label{eq:Du}
\end{equation}
where $\kappa_b$ is the conductivity of the neutral bulk solution,  $\kappa'_s$ is the additional ``surface conductivity" due to excess ions in the EDLs~\cite{bikerman1935,urban1935,deryagin1969,Chu:2007}, and $h$ is a geometrical length scale, such as the channel width or particle size. The competition of surface and bulk conduction in a  microchannel is now well understood for linear response to a small voltage or current~\cite{werner2001,delgado2007,Lyklema:1995}, but recently a surprising nonlinear phenomenon was discovered. 

Mani, Zangle and Santiago showed that, under certain conditions, surface conduction can produce a localized salt concentration gradient propagating through a microchannel, away from a nanochannel junction~\cite{Mani:2009,Zangle:2009}. By deriving a one-dimensional equation for thin EDLs (the ``Simple Model") and applying the method of characteristics, they explained this phenomenon mathematically as shock propagation in the concentration profile, analogous to pressure shocks in gases~\cite{Mani:2009}. The theory was able to predict, for the first time, the propagation of enrichment and depletion shocks in etched glass microchannels on either side of a nanochannel~\cite{Zangle:2009}. It is possible that this phenomenon plays a role in earlier observations of sharp concentration gradients in more complicated microchannel/nanochannel geometries~\cite{zangle:2010rev,zangle:2010,Wang:2005,kim2010,kim2007}.

In this paper, we focus on the new surface-conduction dominated regime and develop a general theory of ``desalination shocks" in complex microstructures. We begin by describing the basic physics of desalination shock propagation in microchannels or porous media. We then develop  general macroscopic transport equations for ions in charged microstructures, which lead to a nonlinear wave equation at constant current. After making the equations dimensionless and identifying the key governing parameters, we study desalination shock propagation in two types of heterogeneous microstructures. First, we analyze slowly varying surface charge and/or channel geometry using perturbation methods, and then we derive intermediate-asymptotic similarity solutions for power-law variations in the channel area. The latter clarify the transition from diffusive scaling ($x\sim \sqrt{t}$) without shocks in a wedge 
to constant-velocity shock propagation in a straight channel ($x \sim t$). Finally, we show that  thin desalination shocks are nonlinearly stable in the absence of fluid flow by reducing the dynamics to a  Laplacian dissolution model.  We conclude by discussing possible applications of our results to microfluidic separations, water desalination, soil decontamination, and energy storage by porous electrodes.

\section{Basic physics of desalination shocks}
\label{basicphysics}
\begin{figure}
  \centering   
  \includegraphics[width=3in]{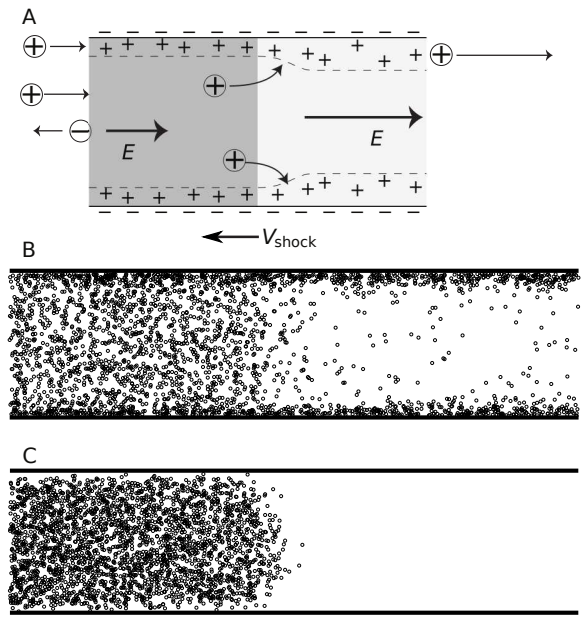}
  \caption{Basic physics of desalination shock propagation. (A) Sketch of ion fluxes in a microchannel or pore with negatively charged walls, as current flows from left to right through a decrease in salt concentration (caused by an electrode or membrane, not shown). In order to avoid low-conductivity region in the center of the channel, the current flows into the electric double layers, where it is carried by positive counter-ions that remain to screen the wall charge. Such ``surface conduction" is driven by the amplified  axial electric field in the depleted region, which also pushes the negative co-ions to the left, thereby sharpening the concentration gradient, leading to a steady shock. These effects are illustrated by snapshots of (B) counterions and (C) co-ions in a Brownian dynamics simulation ~\cite{movie}.}
\label{fig1}
\end{figure}
\begin{figure}
  \centering   
  \includegraphics[width=3in]{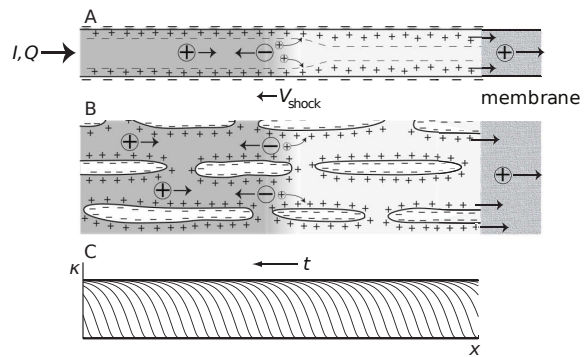}
  \caption{Propagation of desalination shock in a straight microchannel (A) and a homogeneous microporous medium (B). A selective element (membrane) is used at the right-end to trigger an initial depletion (concentration polarization effect), which then propagates in the form of shock through the microstructure. The plot shows the axial profile of the shock uniformly sampled in time (C). For a system at constant current, $I$, and flow rate, $Q$, the shock propagates at a constant speed.}
\label{fig2}
\end{figure}
Consider the passage of current through a microchannel with negatively charged side walls, as shown in Fig.~\ref{fig1}. Suppose that the EDLs are thin and initially play no role in the dynamics. 
An applied voltage drives current from a reservoir on the left to a cation-selective boundary on the right, which only allows cations to pass.  This boundary, shown in Fig.~\ref{fig2}, could represent either  a cation-selective electrodialysis membrane, an electrode where cations are reduced to a neutral species, a negative porous electrode charging capacitively, or one or more nanochannels with over-lapping EDLs. 

In order to maintain electroneutrality as co-ions are expelled, the salt concentration is reduced near the boundary. The ensuing depleted region initially spreads to the left by diffusion. As the bulk conductivity is reduced, however, the axial electric field is amplified (in order to sustain the current) and  acts on the counterions screening the wall charge to drive surface conduction. Regardless of the initial Dukhin number, surface conduction eventually dominates bulk diffusion in carrying the current through the depleted region. Meanwhile, co-ions are driven to the left by the large electric field, thus further enhancing bulk depletion. This nonlinear feedback causes sharpening and propagation of the salt concentration gradient similar to standard shock waves. As shown in Fig.~\ref{fig1}A, current lines are diverted from the bulk solution into the double layers, as they pass through the shock.

In Fig.~\ref{fig1}, we show the results of Brownian dynamics simulations \cite{movie}, which clearly illustrate the physics of shock propagation.  Counterions move from the bulk solution into the double layers in order to carry current around the depleted region behind the shock (Fig.~\ref{fig1}B). Meanwhile, co-ions electromigrate ahead of the shock, and they become fully depleted behind it (Fig.~\ref{fig1}C).  Although molecular simulations allow us to visualize the trajectories of discrete ions, our goal is to elucidate the macroscopic behavior of desalination shocks, so we now turn to continuum models.

\section{ Macroscopic ion transport in microstructures }
The physical arguments above are very general and can be extended to microstructures with other geometries. As shown in Fig.~\ref{fig2}, there is an analogy between macroscopic ion transport in a homogeneous porous medium (Fig.~\ref{fig2}B) and in a microchannel (Fig.~\ref{fig2}A) of suitable thickness, defined below. We begin by considering uniform microstructures, such as constant-height channels and homogeneous porous media (Fig.~\ref{fig2}), and derive general macroscopic transport equations to describe concentration polarization and desalination shocks. We will then extend this model to systems involving geometrical variations, such as variations in porosity or channel cross section.  We simply require that the geometrical and electrochemical properties of the microstructure vary sufficiently slowly to justify a volume averaged model. This basic assumption also underlies formal homogenization analyses~\cite{allaire1992,looker2006,moyne2006,allaire2010,delima2010,schmuck2011} and leads to macroscopic equations for charged porous media of the same general form as we propose below~\cite{schmuck}, but here we will rely on physical arguments without deriving any explicit dependence on the microstructural geometry.

\subsection{ Fluxes and flows}
For simplicity, we use dilute solution theory to model ionic fluxes, but it is straightforward to extend our results by replacing concentrations with activities~\cite{newman2004,large_acis}. Let $c_i$ be the mean volume-averaged concentration of ion species $i$ of charge $q_i$ in the pores (number / pore volume), and $D_i$ be the effective diffusivity within the porous matrix~\cite{newman2004,Torquato:2002}. Conservation of species at the macroscopic continuum level is expressed by the Nernst-Planck equations:
\begin{equation}
\frac{\partial c_i}{\partial t} + \ub\cdot\nabla c_i = \nabla \cdot \left[ D_i \left( \nabla c_i + \frac{q_ic_i}{kT}\nabla \phi \right)\right],
\label{eq:gennp}
\end{equation}
where we have used the Einstein relation to express the mobility of species $i$ as $\nu_i = D_i / kT$ ($k=$ Boltzmann's constant, $T=$ absolute temperature) and $\ub$ is a mean fluid velocity in the pores. As a first approximation, we have neglected dispersion (velocity-dependent effective diffusivity) due to nonuniform convection within the pores~\cite{koch,yaroshchuk}, which is reasonable for thin pores~\cite{dydek}. In addition, we enforce macroscopic incompressibility,
$$
\nabla\cdot\ub = 0,
$$
and postulate linear response to gradients of pressure, potential and concentration at the macroscopic continuum scale,
$$
\ub = - K_H \nabla p - K_E(\{c_i\},\phi) \nabla\phi - \sum_i K_{D,i}(\{c_i\},\phi) \nabla \ln c_i.
$$
The first term is Darcy's law, the second electro-osmotic flow, and the third diffusio-osmotic flow, each of which in principle have tensorial coefficients in an anisotropic medium~\cite{ajdari2002}. The coefficients $K_E$ and $K_{D_i}$ depend on the ionic concentrations, potential and surface charge and could in principle be derived from a microscopic model of intrapore transport or approximations for straight channels with thin double layers.  In our analysis of desalination shocks below, we neglect nonlinearities due to convection to focus on the effects of surface conduction, so we leave the derivation and nonlinear analysis of the full macroscopic transport equations in three dimensions for future work.

\subsection{ Electrostatics} The key source of nonlinearity in our system is the electrostatic coupling between ions and the surface charge of the microstructure. The electrolyte fills a solid matrix of porosity $\epsilon_p$  (pore volume / total volume) and area density $a_p$ (pore area / total volume). The walls of the pores have a fixed charge density $\sigma_s$ (charge / pore area). At the macroscopic continuum scale, {\it the surface charge appears as a fixed background charge density} (charge / pore volume) $\rho_s$ given by
\begin{equation}
\rho_s = \frac{\sigma_s}{h_p} = \frac{\sigma_s a_p}{\epsilon_p},
\label{eq:rhos}
\end{equation} 
where $h_p = \epsilon_p/a_p$ is an effective pore size. In the first step of our derivation, we simply enforce electroneutrality at the macroscopic continuum scale,   
\begin{equation}
\epsilon_p \rho + a_p \sigma_s = 0 \ \ \Rightarrow \ \ \rho = \sum_i q_i c_i = - \rho_s, \label{eq:genen}
\end{equation}
where $\rho$ is the mean ionic charge density, which is equal and opposite to the surface charge density, $\rho_s$. The macroscopic, volume-averaged electroneutrality condition (Eq. \ref{eq:genen}) implicitly determines the mean electrostatic potential in Eq. \ref{eq:gennp}. This approach has also been employed recently to model charge transport in nanochannels~\cite{ramirez2007} and carbon nanotubes\cite{scruggs2009} and can be traced back to early models of ion exchange membranes~\cite{helfferich1962}. 

Let $c = \sum_i |q_i| c_i$ be the total ionic charge (regardless of sign). For $|\rho_s| \ll c$, we recover the standard model for a quasi-neutral bulk electrolyte, which leads to the (ambipolar) diffusion equation for the neutral salt concentration~\cite{newman2004}. In the opposite limit,  $|\rho_s| \approx c$, we recover the standard model for a bulk ion-exchange membrane or solid electrolytes~\cite{rubinstein_book,kornyshev1981,biesheuvel2009}. 
In contrast, our focus is on the intermediate ``leaky membrane" regime, where $|\rho_s| < c$, which generally introduces nonlinearity due to electromigration of the diffuse ionic charge that screens the fixed background charge.

\subsection{Binary electrolyte} 
We consider the canonical unsupported electrolyte: a dilute, asymmetric binary solution ($i=+,-$) with arbitrary ionic charges, $q_\pm = \pm z_\pm e$. In this case, macroscopic transport equations take the form,
\begin{eqnarray}
\frac{\partial c_\pm}{\partial t} + \ub\cdot\nabla c_\pm &=& D_\pm \left[ \nabla^2 c_\pm \pm z_\pm \nabla\cdot\left(c_\pm \nabla \tilde{\phi}\right)\right], 
\label{eq:binnp} \\
0 &=& z_+e c_+ - z_-e c_- + \rho_s,
\label{eq:neutrality}
\end{eqnarray}
where  $\tilde{\phi} = e\phi/kT$ is the dimensionless potential, scaled to the thermal voltage. Without loss of generality, let us assume that the surface charge is negative, $\rho_s<0$, and use Eq. \ref{eq:neutrality} to replace the ion concentrations $c_+$ and $c_-$ with the neutral portion of the salt concentration in the bulk (excluding wall shielding charge)
\begin{equation}
c_b=z_+ c_+ + z_- c_- + \frac{\rho_s}{e} = 2 z_- c_- .
\label{eq:cb}
\end{equation}
In the limit of zero surface charge, this reduces to the total concentration of charges ($c_b \to z_+ c_+ + z_- c_-$) in a neutral electrolyte. In the opposite limit of a fully depleted bulk electrolyte with nonzero surface charge, this quantity vanishes, since only counter-ions remain within the EDLs of the microstructure ($z_+e c_+ \to -\rho_s$). Therefore, the variable $c_b$ measures the amount of ``free conductivity" that can be removed from the microstructure (i.e. contributing to desalination), without disturbing the screening of the fixed surface charge by counter-ions. In terms of these variables, the PDEs can be written in the following form
\begin{eqnarray}
\frac{\partial c_b}{\partial t} + \ub\cdot\nabla c_b &=& \overline{D} \left[ \nabla^2 c_b - \frac{\bar{z}}{e} \nabla\cdot\left(\rho_s\nabla\tilde{\phi}\right) \right], \label{eq:ceq} \\
0 &=& \nabla\cdot\jb,   \label{eq:jeq}
\end{eqnarray}
where $\jb$ is the volume averaged current density (given below); $\overline{D}$ is the ambipolar diffusivity of a binary electrolyte~\cite{newman2004} (see Appendix~\ref{appendixA} for the general form of $\overline{D}$ and $\bar{z}$). 

 It is clear that in this model, any nonlinear response is entirely due to the fixed surface charge, since a {\it linear} convection-diffusion equation for $c_b$ is recovered from Eq. \ref{eq:ceq} if and only if $\rho_s=0$. If any such charge exists in the microstructure, then the second term in Eq. \ref{eq:ceq} survives, and the dynamics of the ionic transport will be coupled to that of the potential $\tilde{\phi}$, which generally satisfies a PDE (Eq. \ref{eq:jeq}) enforcing the conservation of charge. The nonlinearity becomes apparent from the volume-averaged current density in Eq. \ref{eq:jeq}, which takes the form 
\begin{equation}
\frac{e}{kT}(\jb + \rho_s \ub) = - \beta  \nabla \kappa_b - \left[\kappa_b + \frac{\kappa_s}{h_p}\right] \nabla \tilde{\phi},  \label{eq:jb}
\end{equation}
where the second term on the left is the convection of charge; the first term on the right is the diffusion current, controlled by  the parameter
$$
%\beta = \frac{D_+ - D_-}{kT(z_+\nu_+ + z_- \nu_-)},
\beta = \frac{D_+ - D_-}{z_+ D_+ + z_- D_-},
$$
which measures the asymmetry of the electrolyte; the second term on the right hand side of Eq. \ref{eq:jb} is Ohm's law, where the total conductivity is broken into two parts: neutral portion of the bulk, and surface (excess counter-ion) contributions. These are respectively:
\begin{eqnarray}
\kappa_b &=& \frac{ (z_+\nu_+ + z_-\nu_-) e^2 c_b}{2}, %= e \overline{\nu}^e c 
\\
\kappa_s &=& z_+ \nu_+ e |\sigma_s|.% = \nu^e_+ |\sigma_s|
\end{eqnarray}
It is important to stress that what we call $\kappa_s$, which is related to the {\it difference} between co- and counter-ion concentrations (screening the surface charge), is not the same as $\kappa'_s$, the ``surface conductivity". The latter is defined as the excess conductivity due to {\it sum} of co- and counter-ion concentrations in the EDLs relative to the quasi-neutral bulk solution~\cite{bikerman1935,urban1935,deryagin1969,Chu:2007}.

\section{ Conductivity waves at constant current }

%{\bf Uniform current density }

To illustrate the nonlinear dynamics contained in these equations, we consider passing a uniform current density $\jb = j(t) \hat{x}$ and a uniform flow, $\ub =u(t) \hat{x}$ through the porous medium. We solve Eq. \ref{eq:jb} for the electric field and substitute back into Eq. \ref{eq:ceq} to obtain a single, nonlinear PDE for bulk conductivity $\kappa_b(x,t)$:
\begin{equation}
\frac{\partial \kappa_b}{\partial t} + \frac{\partial}{\partial x}\left[ u \kappa_b 
+ \frac{ z_-\nu_-e(\kappa_s/h_p)(j + \rho_s u)}{\kappa_b + \kappa_s/h_p} \right]
= \frac{\partial}{\partial x} \left[ D(\kappa_b) \frac{\partial \kappa_b}{\partial x} \right]
\label{eq:1dp}
\end{equation}
where 
\begin{equation}
D(\kappa_b) = \overline{D}\left(1 - \frac{\bar{z}(D_+-D_-)}{2z_+D_+}\frac{\kappa_s/h_p}{\kappa_b + \kappa_s/h_p}\right).
\end{equation}
This one-dimensional PDE for uniform current is similar to that obtained by Mani, Zangle and Santiago\cite{Mani:2009} in their Simple Model for a flat microchannel with thin double layers. Here, we have generalized the model to porous microstructures, while adding the convective contribution of diffuse charge to the current ($\rho_s u$) as well as the conductivity dependence of the effective diffusivity $D$ for an asymmetric electrolyte interacting with the surface charge.

If the surface effects (the terms with $\kappa_s$) can be neglected, Eq. \ref{eq:1dp} reduces to the classical linear convection-diffusion equation for bulk conductivity. The nonlinear flux, $z_-\nu_-e\kappa_s(j+\rho_su)/(h_p\kappa_b+\kappa_s)$, can be physically interpreted as the advection of the surface charge due to electromigration (as seen in Eq.~\ref{eq:ceq}). Gradients of this flux term are responsible for exchanges between EDL (surface) and the bulk, which are schematically depicted in Fig.~\ref{fig1}.     

Equation \ref{eq:1dp} has the same form as the equations of gas dynamics and shallow water waves~\cite{whitham}, and describes similar nonlinear wave phenomena. In the long time limit in a large system,  convection dominates diffusion and yields a kinematic wave equation of the form,  $c_t + (F(c))_x =0$, which can be solved by the method of characteristics. The basic idea is that initial concentration values propagate with velocity $v_c = F^\prime(c)$ along characteristic lines in space-time. In order to avoid a multi-valued concentration profile, whenever characteristics cross, a discontinuity (or shock) in concentration, $[c]$, is introduced, which moves at the velocity $v_s = [F(c)]/[c]$, where $[F]$ is the jump in flux across the shock. The concentration profile across the shock is  a traveling wave solution, $c(x,t)=f(x-v_s t)$, to the full equation with diffusion. We now apply this kind of analysis to our problem.

\subsection{Dimensionless formulation}
The first step is to define dimensionless variables:
$$
 \tilde{\kappa}=\frac{\kappa_b}{\kappa_{b\infty}},\,\,\, \tilde{x}=\frac{x}{\overline{D}}\frac{z_-\nu_-ej}{\kappa_{b\infty}},\,\,\, \tilde{t}=\frac{t}{\overline{D}}\left(\frac{z_-\nu_-ej}{\kappa_{b\infty}}\right)^2,
$$ 
where $\kappa_{b\infty}$ is the reference bulk conductivity (typically in a reservoir connecting to the microstructure). Space and time coordinates are nondimensionalized using diffusive scaling together with characteristic electrodiffusion velocity, $z_-\nu_-ej/\kappa_{b\infty}$. With these definitions, Eq. (\ref{eq:1dp}), takes the following dimensionless form 
\begin{equation}
 \frac{\partial \tilde{\kappa}}{\partial \tilde{t}}+\frac{\partial}{\partial \tilde{x}}\left(\tilde{u}\tilde{\kappa}+\frac{\Du}{\tilde{\kappa}+\Du}\right)=\frac{\partial^2 \tilde{\kappa}}{\partial \tilde{x}^2},
  \label{nondimensional:1}
\end{equation} 
where, for simplicity, we have neglected asymmetric diffusion ($D = \overline{D}$) and the convection of diffuse charge ($|\rho_s u| \ll |j|$). In this equation, two fundamental dimensionless groups appear. The first parameter, 
\begin{equation}
\tilde{u}=\frac{u\kappa_{b\infty}}{z_-\nu_-ej},
\end{equation}
is the ratio of the mean fluid velocity, $u$, to the electrodiffusion velocity, $z_-\nu_-ej/\kappa_{b\infty}$. This parameter affects the shock propagation velocity (essentially a Galilean transformation), but not its dynamics. The second, more important, parameter in Eq. \ref{nondimensional:1} is a dimensionless surface charge, 
\begin{equation}
\Du=\frac{\kappa_s}{h_p\kappa_{b\infty}} = 
\frac{|\sigma_s|} {h_p\left( 1 + \frac{z_- \nu_-}{z_+\nu_+}\right) z_- e c_{\infty-}}.
\label{Dukhin}
\end{equation} 
With our notation the dimensionless parameter $\Du$ in Eq. \ref{Dukhin} resembles the Dukhin number, $\Dukhin$, in Eq. \ref{eq:Du}, but, as discussed above, they are not the same ($\kappa_s\ne\kappa_s'$). 
 
For typical concentrations in aqueous solutions, $\Du$ is very small for microstructures ($h_p\sim1\mu$m), suggesting that the nonlinear term in Eq. \ref{nondimensional:1} can be neglected. One mechanism that can activate the nonlinear term (and produce shocks) is to locally decrease $\tilde{\kappa}$ to very small values of order $\Du$. This is the crucial role that the selective boundary (e.g. the membrane in Fig.~\ref{fig2}A) plays in these systems. 

As the shock propagates, it leaves behind a region with orders of magnitude lower salt concentration. In other words, propagation of the shock acts to {\it desalinate the bulk electrolyte}. In the next two sections, we analyze the dynamics of desalination shocks in systems with non-uniform geometries.

\section{Weakly varying microstructures} 
\label{sec:weakly}
\begin{figure}
  \centering   
  \includegraphics[width=3in]{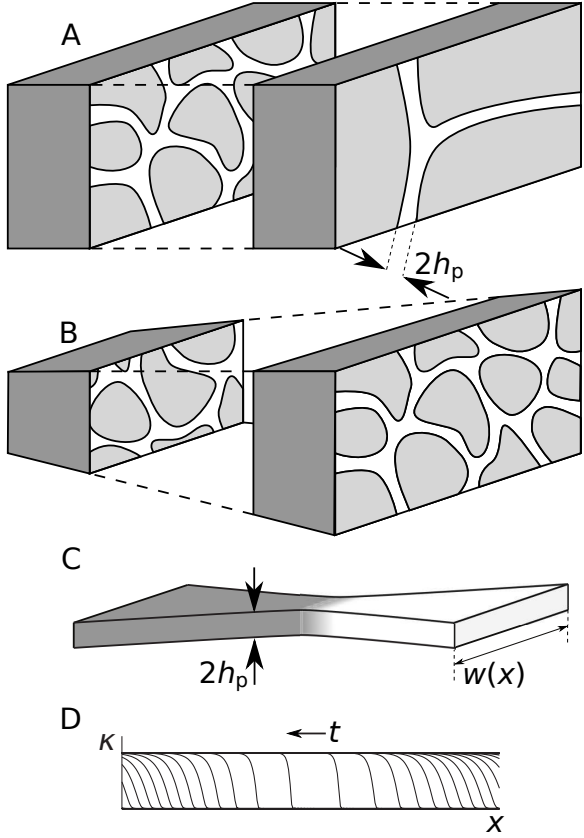}
  \caption{Weakly varying microstructures with constant pore size, $h_p$. Schematics include a microstructure with variable porosity, $\epsilon_p$, and area density, $a_p$, but fixed $h_p=\epsilon_p/a_p$ (A), a homogeneous microstructure with variation in the macroscopic geometry (B), and a fabricated microchannel with variable-width (C). Propagation of a depletion shock through the converging-diverging channel under the constant current and flow rate condition is shown (D). The plots are sampled uniformly in time.}
\label{fig3}
\end{figure}

\begin{figure}
  \centering   
  \includegraphics[width=3in]{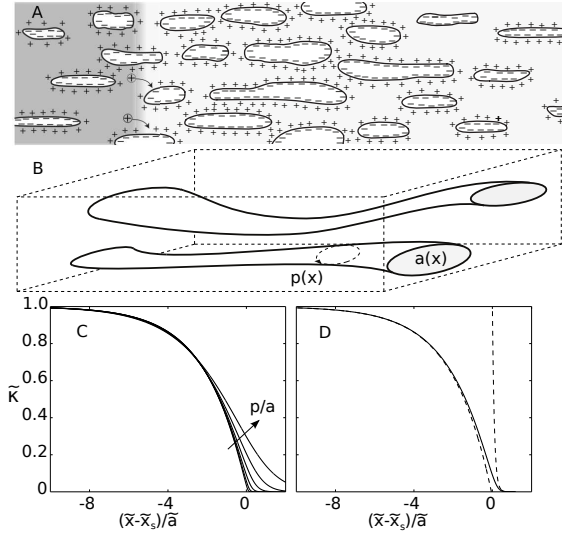}
  \caption{Microstructures with varying pore size, $h_p$. Figure shows example of a porous medium (A) and a microtube (B). Plots show the bulk conductivity versus axial length across a desalination shock (C) for $\tilde{\kappa}_d=\Du=0.025$. Minimum $\tilde{p}/\tilde{a}$ is 0.25 and is doubled for each subsequent plot up to $\tilde{p}/\tilde{a}=8$. Figure D shows the plot for $\tilde{p}/\tilde{a}=1$ with the dashed lines representing the left and right asymptotic curves.}
\label{fig4}
\end{figure}
Figures~\ref{fig3} and \ref{fig4} show examples of structures involving variation of porosity, pore-size, and macroscopic geometry. The analysis presented in the previous section can be easily extended to these structures. Our analysis only requires that the microstructure properties vary slowly enough to allow a local volume averaged theory. While the general derivation is presented in Appendix~\ref{appendixA}, we here continue to focus on the simplified quasi-one-dimensional systems and study the response of desalination shocks to structural inhomogeneities. Under such conditions the modified form of Eq. \ref{eq:1dp} can be obtained by simply scaling all the flux and rate terms with appropriate local volume and/or area measures (see below).
 
\subsection{Structures with constant pore size}
We first consider weakly-variable microstructures with constant pore size. In other words, in these structures, porosity and area-density vary proportionally. With a constant surface charge, these structures have a uniform background charge density, $\rho_s$ (see Eq. \ref{eq:rhos}). Figure~\ref{fig3} shows examples of such structures, in which the net local volume changes as a function of axial coordinate. In Fig.~\ref{fig3}A the net cross-sectional area (different from area-density) is proportional to local porosity, $\epsilon_p$; in Fig.~\ref{fig3}B it is proportional to local macroscopic area; and in Fig.~\ref{fig3}C it is proportional to microchannel width, $w$. These parameters essentially play the same role in modifying the dynamics of desalination shocks by scaling the fluxes in the conservation laws. For example, for the case of the variable-width microchannel, Eq. \ref{eq:1dp} (again, setting $\rho_s u=0$ and $D=\overline{D}$) will be modified to:
\begin{equation}
\frac{\partial}{\partial t}(w\kappa_b) + \frac{\partial}{\partial x}\left[ uw \kappa_b 
+ \frac{ z_-\nu_-ejw\kappa_s/h_p}{\kappa_b + \kappa_s/h_p} \right]
= \frac{\partial}{\partial x} \left( w\overline{D} \frac{\partial \kappa_b}{\partial x} \right),
\label{eq:conductivity_w}
\end{equation}
where the ``volume averaged'' quantities, $\kappa_b$, $u$, and $j$ are effectively the height-averaged quantities , and the equivalent pore-size, $h_p$, is half the channel height. To be able to neglect the transverse fluxes and reduce the system to one-dimensional PDE we need the macroscopic geometry to vary with small slope ($dw/dx\ll 1$), which is a standard assumption of lubrication theory. The gradually varying assumption imposes an additional condition which physically means that macroscopic properties do not change much over the axial thickness of the shock. We use $u_0$, and $j_0$, evaluated at $x_0$ (shock location at $t_0=0$) to nondimensionalize Eq. \ref{eq:conductivity_w}. $w$ can be nondimensionalized using $w_0$. Noting that $uw$ and $jw$ are constant in $x$ due to conservation of mass and current, Eq. \ref{eq:conductivity_w} can be nondimensionalized to  
\begin{equation}
 \frac{\partial}{\partial \tilde{t}}(\tilde{w}\tilde{\kappa})+\frac{\partial}{\partial \tilde{x}}\left(\tilde{u}\tilde{\kappa}+\frac{\Du}{\tilde{\kappa}+\Du}\right)=\frac{\partial}{\partial \tilde{x}}\left[\tilde{w}\frac{\partial \tilde{\kappa}}{\partial \tilde{x}}\right].
 \label{nondimensional:2}
\end{equation} 
One can verify that Eq. \ref{nondimensional:2} is also applicable to the case of porous media. In that case $\tilde{w}$ would be the nondimensional net cross sectional area. In this formulation $\tilde{u}$ and $\tilde{\rho}_s$ are the nondimensional constant parameters, and $\tilde{w}$=$\tilde{w}(\tilde{x})$ is a known function.   
Equation~\ref{nondimensional:2} has the trivial boundary condition of $\tilde{\kappa}_{-\infty}=1$. We also use a Dirichlet boundary condition of $\tilde{\kappa}(\tilde{x}=0)=\tilde{\kappa}_d=O(\Du)$, which represents a depletion boundary, initiated by a selective element next to the channel. We seek a solution of the form 
\begin{equation}
\tilde{\kappa}(\tilde{x},\tilde{t})=f(\eta)=f\left(\frac{\tilde{x}-\tilde{x}_s(\tilde{t})}{\tilde{l}_s(\tilde{t})}\right),
\label{eq:steady_form}
\end{equation}
where $\tilde{x}_s$ represents the shock location and $\tilde{l}_s$ is the shock length or axial thickness. The profile of $\tilde{f}$ satisfies an ODE, yet to be obtained. Since this profile should look like a shock, we have $f(\eta\ll-1)\simeq 1$ ,and $f(\eta\gg1)\simeq \tilde{\kappa}_d$. We propose a solution for $\tilde{x}_s(\tilde{t})$ and  $\tilde{l}_s(\tilde{t})$ by speculating that the local shock length is proportional to the local channel width and its speed is inversely proportional to the width:
\begin{equation}
\frac{d\tilde{x}_s}{d\tilde{t}}=\frac{\tilde{v}}{\tilde{w}\left(\tilde{x}_s(\tilde{t})\right)},\,\,\,\tilde{l}_s(\tilde{t})= \tilde{w}\left(\tilde{x}_s(\tilde{t})\right),
\label{eq:steady_scaling}
\end{equation}
where $\tilde{v}$ is the dimensionless shock speed at $x=x_0$. By substituting Eq. \ref{eq:steady_scaling} into Eq. \ref{eq:steady_form}, then into the governing equation (Eq. \ref{nondimensional:2}), and ignoring variations of $\tilde{w}$ over the shock thickness we obtain the following ODE for $f$.
\begin{equation}
{\left[(\tilde{u}-\tilde{v})f+\frac{\Du}{f+\Du}\right]}^\prime=f^{\prime\prime}.
\label{eq:steady_ode}
\end{equation}
To compute the constant $\tilde{v}$, we can integrate Eq. \ref{eq:steady_ode} from $-\infty$ to $+\infty$ and use the boundary conditions. Since $f^\prime=0$ in the limits, we obtain
\begin{eqnarray}
\tilde{v}=\tilde{u}-\frac{\Du}{\tilde{\kappa}_d + \Du}+O(\Du).
\label{eq:shock_vel_condition}
\end{eqnarray}
 Note that shock propagation would be possible only for negative $\tilde{v}$. This can be typically accommodated only if sufficient depletion is introduced at the boundary by $\tilde{\kappa}_d=O(\Du)$ (also needs $\tilde{u}<1$).

Substituting Eq. \ref{eq:shock_vel_condition} into Eq. \ref{eq:steady_scaling} and rewriting in the dimensional form reveals that for strong shocks (i.e. $\tilde{\kappa}_d\sim\Du\ll 1$) the {\it local} shock velocity relative to the {\it local} flow is
\begin{equation}
\frac{dx_s}{dt}-u(x)=-\left(\frac{z_-\nu_-ej(x)}{\kappa_{b\infty}}\right)\frac{1}{1+h_p\kappa_d/\kappa_s}.
\label{eq:shockspeed}
\end{equation}
The right-hand-side of Eq. \ref{eq:shockspeed} is the electrodiffusion velocity in the enriched side of the shock scaled by a rational function of the surface to bulk conduction in the depleted side. As physically expected, in the limit of perfect desalination, $\tilde{\kappa}_d=0$, the relative shock velocity will be identical to the coion electromigration velocity.   

Integrating Eq. \ref{eq:steady_scaling} yields
\begin{equation}
\int \tilde{w}(\tilde{x}_s)d\tilde{x}_s=\tilde{v}\tilde{t},
\end{equation}
which indicates that {\it the rate of sweeping the volume of the channel by the shock is constant}. This also makes sense from the global conservation law point of view: Very far from the shock, at the channel boundaries, the flux term, $\tilde{u}\tilde{\kappa}+(\Du)/(\tilde{\kappa}+\Du)$ (see Eq.~\ref{nondimensional:2}), does not change with time and the diffusion flux is negligible. From global conservation, the depletion of ions inside should balance the difference of the fluxes at the boundaries. Therefore, the depletion rate should be constant, implying the rate of sweeping the volume by the shock should be constant.

\subsection{Microstructures with variable pore size}
This powerful observation can be generalized to more complicated microstructures such as the ones shown in Fig.~\ref{fig4}. In this case, as shown in Fig.~\ref{fig4}A, we deal with a microstructure with gradually varying porosity, $\epsilon_p$ and surface density, $a_p$, independent of each other. Equivalently we also can consider microtubal structures (see Fig.~\ref{fig4}B) with gradual variation in cross-sectional area, $a(x)$, and cross-sectional perimeter, $p(x)$. Under our simplifying assumption of quasi-one-dimensional systems, $\epsilon_p$ in the microporous media plays the equivalent role of $a(x)$ in microtubal structures; they both scale the bulk quantities. In addition, the role of $a_p$ in porous media is analogous to the role of $p(x)$ in microtubes; they both scale the surface quantities. For the case of microtubes, the modified governing equation is  
\begin{equation}
\frac{\partial}{\partial t}(a\kappa_b)+\frac{\partial}{\partial x}\left(ua\kappa_b+\frac{z_-\nu_-eja\kappa_s/h_p}{\kappa_b+\kappa_s/h_p}\right)=\frac{\partial}{\partial x}\left[a\overline{D}\frac{\partial \kappa_b}{\partial x}\right],
\label{eq:general}
\end{equation}
where the ``volume averaged'' quantities, $\kappa_b$, $u$, and $j$ are effectively the cross-sectional averaged quantities for the case of a microtube. The equivalent pore-size, $h_p$, is a/p. Equation~\ref{eq:general} is very similar to Eq. \ref{eq:conductivity_w} with the exception that now $h_p$ is not a constant and is equal to $a(x)/p(x)$. Again, as a shock propagates, it sweeps the net available volume of the structure at a constant rate independent of complexities of $a(x)$ and $p(x)$. 

For the case of constant-$h_p$ we showed that the shock axial extent would be proportional to local area of the channel. For general $a$ and $p$ however, the evolution of shock length is not as simple. It turns out that even a solution with the form presented by Eq. \ref{eq:steady_form} is not valid any more. In this general case, different regions of the shock can scale differently. We here only report the analytical solution to the shock profile and refer the reader to Appendix~\ref{appendixB} for details of the derivation. One can show that $\tilde{\kappa}$ changes as a function of axial coordinate  according to the following relation (see Fig.~\ref{fig4}C)
\begin{equation}
\left(\frac{\Du}{\tilde{\kappa}_d+\Du}\right)\frac{\tilde{x}-\tilde{x}_s}{\tilde{a}}=\text{ln}(1-\tilde{\kappa}) - (\tilde{\kappa}_d+\Du) \frac{\tilde{p}}{\tilde{a}} \ln\left(\tilde{\kappa}-\tilde{\kappa}_d \frac{\tilde{p}}{\tilde{a}}\right),
\label{eq:shockprofile}
\end{equation}
where $\tilde{\kappa}_d$ and $\Du$ are constants: $\tilde{\kappa}_d$ is the dimensionless bulk conductivity at the depletion boundary, and $\Du$ is  $\kappa_s/ h_p(x_0)\kappa_{b\infty}$. $\tilde{a}$ and $\tilde{p}$ are gradually varying local area and perimeter nondimensionalized by their reference values at $x_0$. 

With $\tilde{a}$ in the denominator of the left-hand-side, this format indicates that the shock axial thickness scales with local $\tilde{a}$ (as seen previously), but its shape depends on parameter $\tilde{p}/\tilde{a}$. The right-hand-side of Eq.~\ref{eq:shockprofile} involves two terms: The first term, $\text{ln}(1-\tilde{\kappa})$, is dominant in high concentration region ($\tilde{\kappa}\gg \Du$); the second term, which involves $\tilde{p}/\tilde{a}$ as a parameter, is of order $O(\Du)$ and is dominant in low concentration zone of the shock. A plot of the shock profile together with these two asymptotic profiles are presented in Fig.~\ref{fig4}C. As mentioned before, one can observe that the shock profile is independent of convection parameter $\tilde{u}$.

From physical standpoint it is worth noting that the asymptotic profile of the shock on the high-concentration side,
\begin{equation}
\tilde{\kappa} \sim 1 - \exp \left[ \left(\frac{\Du}{\tilde{\kappa}_d+\Du}\right)\frac{\tilde{x}-\tilde{x}_s}{\tilde{a}}\right]
\label{eq:shockprofile2}
\end{equation}
is governed by axial diffusion and a low-concentration boundary condition, moving relative to the bulk flow. The nonlinear transport associated with surface conductivity is negligible through this high-conductivity zone, although it plays a role in determining the velocity. The same propagating exponential concentration profile of Eq. \ref{eq:shockprofile2} also arises in other situations, such as dendritic electrodeposition~\cite{Bazant:1995,Leger:1998}, where counter-ions are removed by convection-diffusion-reaction processes at the dendrite tips~\cite{huth1995}, rather than by surface conduction.

\section{Similarity solutions for power-law growth of area}
\label{sec:similarity}
\subsection{Intermediate asymptotics}
\begin{figure}
  \centering   
  \includegraphics[width=3in]{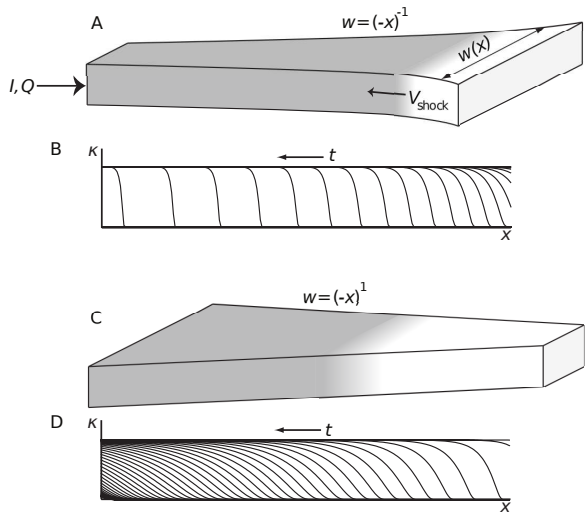}
  \caption{Schematics of desalination shock propagation in a contracting microchannel ($\gamma=-1$) is shown in A. Profiles of the shock at different stages indicate that as the shock reaches the narrower regions of the microchannel it gains speed and adopts a sharper axial profile (B). Schematics of propagation in a linearly expanding channel is shown in C. Time-series of the axial profiles indicate that the shock slows down and becomes diffuse towards the end of the channel (D).} 
\label{fig5}
\end{figure} 
 In this section we consider the constant-pore-size structures again, but with power law growth of their area, $\tilde{w}={\left(-\tilde{x}\right)}^\gamma$, as shown in Fig.~\ref{fig5}. Note that in our notation $\tilde{w}$ represents nondimensional cross-sectional area (or equivalently channel width or porosity) for a microstructure with constant pore size. In this section, variation of $\tilde{w}$ is not necessarily negligible over the shock axial extent. We are interested in solutions to Eq. \ref{nondimensional:2} at large enough times to approach a self-similar form. Such ``intermediate asymptotic" solutions~\cite{barenblatt96} with power-law monomial scalings are expected based on dimensional analysis~\cite{barenblatt87}, due to the lack of any natural length scale in the problem. We seek asymptotic solutions of the form
\begin{equation}
\tilde{\kappa}=f(\eta)=f(\frac{\tilde{x}+C\tilde{t}^\alpha}{\tilde{t}^\beta}).
\label{eq:similarity_variable}
\end{equation}
which describe features that advect with the scaling $\tilde{t}^\alpha$ as they enlarge (thicken) with the scaling $\tilde{t}^\beta$. Our objective is to find $\alpha$ and $\beta$ as functions of $\gamma$. Note that $\alpha>\beta$ would indicate a shock-like solution where propagation is faster than growth of the structure; $\alpha<\beta$ indicates a diffusion-like spreading, in which advection is not observable due to the fast growth of the structure itself. Substituting this solution into Eq. \ref{nondimensional:2}, and simplifying results in
\begin{equation}
\left[\frac{1}{\tilde{t}}{\left(C\tilde{t}^\alpha-\eta\tilde{t}^\beta\right)}^\gamma\left(C\alpha\tilde{t}^\alpha-\beta\eta\tilde{t}^\beta\right)+\gamma{\left(C\tilde{t}^\alpha-\eta\tilde{t}^\beta\right)}^{\gamma-1}\right]f^\prime
+{\left(\tilde{u}f+\frac{\Du}{f+\Du}\right)}^\prime=\frac{{\left(C\tilde{t}^\alpha-\eta\tilde{t}^\beta\right)}^\gamma}{\tilde{t}^\beta}f^{\prime\prime}.
\label{eq:similarity_check}
\end{equation}
In the large $\tilde{t}$ limit appropriately selected $\alpha$ and $\beta$ would reduce this equation to an ODE for $f$. Table~\ref{table1} summarizes the resulting $\alpha$ and $\beta$ for different $\gamma$ scenarios. Following Bazant and Stone~\cite{bazant2000}, one can systematically check that these are the only scalings that satisfy the boundary conditions, but we omit such mathematical details here. Note that for the case $\gamma<-1$, the total volume of the medium is finite, and an intermediate asymptotic limit does not exist.

\begin{table}
\caption{Scaling of desalination shock advancement and thickening with time for a microchannel with power law growth of width. $\gamma$ is power of growth of channel width with axial coordinate, $w=(-x)^\gamma$; the shock location is assumed to advance as $x_s\sim t^\alpha$; and the shock axial thickness grows/shrinks as $l_s\sim t^\beta$.
}
\begin{center}          
%\begin{minipage}{3in}
\begin{tabular}{|c|c|c|c|c|}
\hline 
$\gamma$    &  -1 & $(-1,1)$ & 1 &$(1,\infty)$ \\ \hline
 $\alpha$ &  exponential & $\frac{1}{\gamma+1}$ & $\frac{1}{2}$ & --- \\ \hline
$\beta$ &  exponential & $\frac{\gamma}{\gamma+1}$ &  $\frac{1}{2}$ & $\frac{1}{2}$ \\ \hline
description & shock & shock & shock/diffuse & diffuse \\
\hline
\end{tabular}
\label{table1}
%\end{minipage}
\end{center}
\end{table} 
 
\subsection{Exponential shock propagation}
In the singular case of $\gamma=-1$ the formal values of $\alpha$ and $\beta$ are infinite. Under this condition the correct solution would be shock propagation with exponential acceleration in time and the correct similarity variable is $\eta=(\tilde{x}+e^{\alpha'\tilde{t}})/e^{-\alpha'\tilde{t}}$. In the limit of large $\tilde{t}$ the PDE can be transformed to the following ODE:
\begin{equation}
{\left[(\tilde{u}+\alpha')f+\frac{\Du}{f+\Du}\right]}^\prime=f^{\prime\prime}.
\label{eq:n3}
\end{equation}
Similar to what observed in  Eq. \ref{eq:steady_ode}, the value of $\alpha'$ can be obtained by integrating the above equation from $-\infty$ to $+\infty$ and using the boundary conditions. 
\begin{eqnarray}
\alpha' =\frac{1}{1+\tilde{\kappa}_d/\Du}-\tilde{u}+O(\Du).
\label{eq:shock_vel_condition_n3}
\end{eqnarray} 
The parameter $\alpha'$ can be interpreted as the inverse of the time scale for exponential propagation and spreading of the concentration profile.

\subsection{Power-law shock propagation}
For $-1<\gamma<1$ the problem has a power law similarity solution with $\alpha=1/(\gamma+1)$ and $\beta=\gamma/(\gamma+1)$. Note that for this range $\alpha>\beta$ and thus the solution indicates shock propagation. In the limit of large $\tilde{t}$ Eq. \ref{eq:similarity_check} reduces to the following ODE:
\begin{equation}
{\left[\left(\tilde{u}+\frac{C^{\gamma+1}}{\gamma+1}\right)f+\frac{\Du}{f+\Du}\right] }^\prime=C^\gamma f^{\prime\prime}.
\label{eq:n1}
\end{equation} 
Interestingly, in the limit of $\gamma=1$ this solution leads to $\alpha=\beta=1/2$, which represents the onset of transition towards a diffusive propagation. 

\subsection{Diffusive shock propagation in a wedge (critical case)}
The case of $\gamma=1$ represents a structure with linear growth of area. A practical example, is a wedge-like channel whose width grows with constant slope as shown in Fig.~\ref{fig5}C.  After the case of a straight channel ($\gamma=0$), this case maybe the most relevant for lab-on-a-chip systems. Note that for $\gamma=1$ equations can be represented in cylindrical coordinates (with $\tilde{x}$ interpreted as radius); the lubrication theory assumption ($dw/dx\ll1$) is not necessary to enable reduction of the system to one-dimensional PDE. Therefore, the wedge angle can be any number from $0$ to $2\pi$.

For $\gamma=1$ the similarity variable reduces to $\eta=\tilde{x}/\sqrt{\tilde{t}}$,  which shows diffusive scaling in time.  Equation~\ref{eq:similarity_check} reduces to 
\begin{equation}
-\left(\frac{\eta}{2}+\frac{1+\tilde{u}}{\eta}\right)f^\prime-\frac{1}{\eta}\left(\frac{\Du}{f+\Du}\right)^\prime=f^{\prime\prime},
\end{equation}  
but there is still some effect of surface conduction, measured by $\Du$.

\subsection{Linear diffusion (no shocks)}
For all values of $\gamma>1$ the similarity variable will also be $\eta=\tilde{x}/\sqrt{\tilde{t}}$ and Eq. \ref{eq:similarity_check} reduces to the following ODE, which corresponds to linear diffusion:
\begin{equation}
-\left(\frac{\eta}{2}+\frac{\gamma}{\eta}\right)f^\prime=f^{\prime\prime}.
\label{eq:n2}
\end{equation}
This ODE is valid for large $\tilde{t}$, when the advective flux term in Eq. \ref{eq:similarity_check} becomes negligible compared to other terms. Note that there is no longer any effect of surface conduction ($\Du$) on the intermediate asymptotic similarity solution.

In the case that variation of $\tilde{w}$ is due to change in the macroscopic geometry, such as in microchannels, for very large $\tilde{t}$ the diffusive front may reach locations of the channel with large $dw/dx$ and the lubrication theory assumption may not be valid any more. As a result Eq. \ref{eq:n2} will be valid for these structures only for a range in time described by:
\begin{equation}
1 \ll \tilde{t}^{\frac{\gamma-1}{2}}\ll \frac{D\kappa_{b\infty}}{\gamma w_0z_-\nu_-ej_0}.
\end{equation}   
For durations much larger than the upper bound, the channel span would have a fast growth, $dw/dx\gg1$. In this range, the channel maybe approximated by a 180-degree wedge and propagation can be modeled by the axisymmetric case ($\gamma=1$). 

\subsection{Transients to similarity solutions}
\begin{figure}
  \centering   
  \includegraphics[width=3.2in]{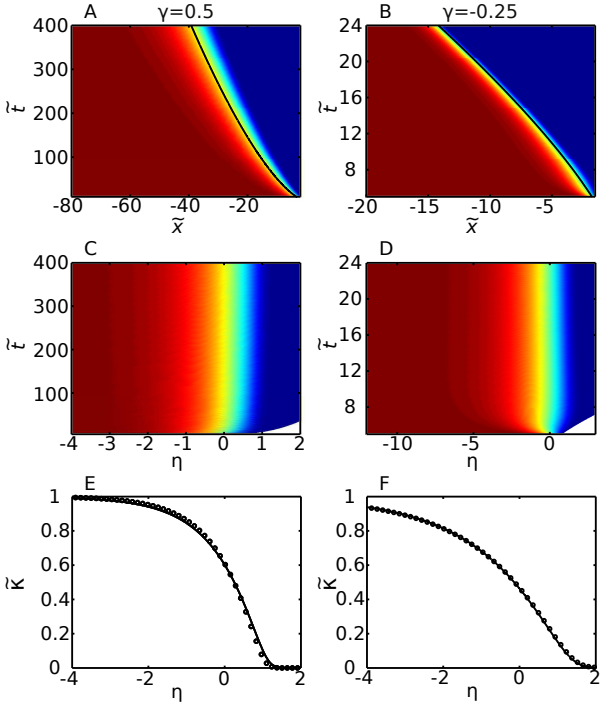}
  \caption{Spatio-temporal evolution of the desalination shock for an expanding microchannel with $\gamma$=0.5 (A) and a contracting microchannel with $\gamma$=-0.25 (B). For both channels $\tilde{u}=0.5$ and $\Du=0.1$. The black line represents $\tilde{x}=-c\tilde{t}^{1/(\gamma+1)}$, where $c$ is 0.72 in A and 0.21 in B. When the data is plotted against $\eta=\left(\tilde{x}+c\tilde{t}^{1/(\gamma+1)}\right)/\tilde{t}^{\gamma/(\gamma+1)}$, the temporal evolution collapses to a single profile after sufficient time (C,D). Concentration profile at the last time instant (symbol) is compared to the asymptotic profile from solution of Eq. \ref{eq:n1} (E,F).} 
\label{fig6}
\end{figure} 
Figure~\ref{fig6} shows a comparison of numerical solutions of  the full model, Eq. \ref{nondimensional:2}, with our similarity solutions for an expanding channel with $\gamma=0.5$ and a converging channel with $\gamma=-0.25$. The spatio temporal plots in Fig.~\ref{fig6}A and Fig.~\ref{fig6}B show that the shock decelerates and becomes smeared by diffusion in the expanding case; conversely in the converging channel, the shock sharpens and accelerates. Representation of these plots in terms of the similarity variable, $\eta$, shows that after a short (dimensionless) transient time the contours collapse into a single self-similar profile, as in other problems of intermediate asymptotics\cite{barenblatt96}. Comparison with the concentration profile obtained from the full model demonstrates the satisfactory accuracy of the similarity solutions.

\section{Nonlinear stability of desalination shocks}
%\label{sec:stability}

So far, we have focused on one-dimensional shock profiles, but these are not special cases of the macroscopic (volume averaged) nonlinear dynamics. Instead, we expect these solutions to be stable attractors, in the sense of intermediate asymptotics~\cite{barenblatt96}, at least in the absence of flow or sudden property changes ($\sigma_s$ and $h_p$).  To make this case, we consider a  ``thin shock", whose thickness is much smaller than its local radius of curvature, under conditions of strong depletion ($\tilde{\kappa}_d=0$). In this limit, the desalinated side contains only surface conductivity, thus the Ohm's law in this region would be of the form: $
\jb=-(\kappa_s/h_p)\nabla\phi$. Conservation of charge then implies that the potential is harmonic, away from the shock:
\begin{equation}
\nabla^2\phi=0  \mbox{ for } {\bf x} \in \Omega(t),
\end{equation}
where $\Omega(t)$ represents the desalinated domain. The region  ahead of the shock has much larger conductivity than the desalinated region, so most of the voltage drop is sustained in the desalinated region. In this limit, the variation of potential outside of $\Omega$ can be neglected compared to the scale of potential-variation inside $\Omega$:
\begin{equation}
\phi=0  \mbox{ for } {\bf x} \in \partial\Omega(t), 
\end{equation}
where $\partial\Omega(t)$ is the boundary specified by the shock location, $x=x_s$. 

\begin{figure}
  \centering   
  \includegraphics[width=3in]{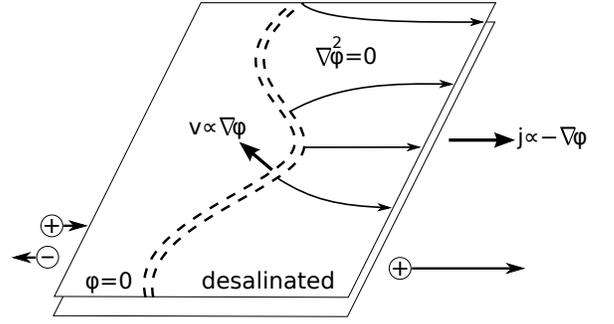}
  \caption{Stability and nonlinear evolution of thin desalination shocks in higher dimensions in the absence of flow. The potential is approximately harmonic in the desalinated region behind the shock and constant in the high-conductivity region ahead of the shock, and the shock moves in proportion to the local electric field, which drives co-ion removal. This problem is mathematically equivalent to Laplacian dissolution~\cite{bazant2006}, a well known stable process that leads to smooth interfaces from arbitrary initial conditions.} 
\label{fig7}
\end{figure}

Next, we obtain an equation for boundary-movement in terms of potential. As described by Eq.~\ref{eq:shockspeed}, in the limit of perfect desalination ($\kappa_d=0$), the shock velocity is same as local electrodiffusion velocity of the coion species:
$$
{\bf v}_s=-\frac{z_-\nu_-e}{\kappa_{b\infty}}{\bf j}.
$$
Since ${\bf j}$ is continuous across the shock, it can be written using the Ohm's law evaluated at the desalinated side of the boundary:
\begin{equation}
{\bf v}_s=+\left(\frac{z_-\nu_-e\kappa_s}{h_p\kappa_{b\infty}}\right)\nabla\phi.
\end{equation}
As shown in Fig.~\ref{fig7}, the resulting model is mathematically equivalent to the well-known problem of Laplacian growth, where an equipotential boundary climbs the normal gradient of a harmonic function, only here it is {\it time reversed}, i.e. the boundary propagates away from the harmonic domain. In two dimensions, Laplacian growth can be solved using time-dependent conformal maps, and it is known to be unstable when the boundary advances into the harmonic domain, leading to cusp-like singularities in finite time~\cite{howison1992}. Physically, this situation is like dendritic electrodeposition or viscous fingering, where air displaces water in a Hele-Shaw cell (without surface tension)~\cite{bensimon1986}.
In contrast, thin desalination shocks evolve by the time-reversed process, which is extremely stable and tends to smooth, symmetric shapes. Physically, desalination shock dynamics resemble  water displacing air in a Hele-Shaw cell or porous medium, or (quasi-steady) diffusion-limited dissolution of a porous solid. Dissolution fronts are often so stable that they can maintain a macroscopic planar shape, even when passing through a highly disordered medium~\cite{leger1999,harris2010}. For several classes of analytical solutions of the time-reversed Laplacian growth see Ref.~\cite{bazant2006}. 

This insight justifies {\it a posteriori} a key assumption in our similarity solutions above. It also shows that they represent universally long-time limits for broad classes of initial conditions. We leave for future work questions of how fluid flow and shock structure might affect this picture. Besides microscopic hydrodynamic instabilities within the microchannels noted above, we cannot rule out the possibility of macroscopic instabilities of desalination shocks, e.g. with misaligned fluid flow and electrical current.

\section{Conclusion and Outlook}
\label{sec:conclusion}
In summary, we have developed a general theory of ion transport in microchannels and porous media, focusing on the new nonlinear regime where surface conduction dominates convection in competing with bulk diffusion. For slowly varying microstructures, the equations support propagating shocks, as well as similarity solutions with power-law scalings. Even in the presence of microscopic inhomogeneities, we expect that these solutions are stable attractors of the nonlinear dynamics.  The multidimensional problem is more complicated, especially in situations where the current is misaligned with the fluid velocity. We believe this system  provides many promising directions for research in applied mathematics.

As suggested by our choice of nomenclature, a natural application of our theory would be to water purification and desalination using porous media and membranes. The basic idea is to extract fresh water continuously from the region behind a steady desalination shock.  Our group is currently investigating this concept~\cite{patent}, and the results will be reported elsewhere. 

Desalination shocks could also be used to enhance the electrokinetic decontamination of  microfluidic devices and porous rocks, clays or soils~\cite{shapiro1993,probstein1993}.  The propagation of  a desalination shock would push co-ionic impurities ahead of the shock, while counterionic impurities would be swept behind the shock by the large electric field.  This effect, driven by surface conduction, promotes the sharpening of the particle profile by electromigration~\cite{bharadwaj2005}, which can also lead to shocks when the particles significantly alter the conductivity ~\cite{ghosal2010}.

Our theoretical results could also be applied to DC electro-osmotic pumps, which employ electro-osmotic flow in porous glass frits~\cite{yao2003b,brask2005,laser2004}. Strickland et al.~\cite{strickland2010} and Suss et al.~\cite{suss} have recently found that concentration polarization can be a key factor in the pump performance, but current theories do not account for the formation of concentration gradients or surface conduction. 

Our results may also find applications in micro/nanofluidic systems. We have shown that varying the cross-sectional area,  perimeter and/or surface charge of a microchannel provides robust means to control the nonlinear dynamics of transport. In parameter regimes where surface conduction is important, this capability may be useful in microfluidic devices for biological sample pre-concentration~\cite{Wang:2005} and seawater desalination~\cite{kim2010} consisting of microchannel/nanochannel junctions. During normal operation, complex electrokinetic instabilities have been observed~\cite{kim2007} and, together with fast pressure-driven flows~\cite{kim2010}, electrohydrodynamic phenomena may dominate any effects of surface conduction. Geometrical optimization of microchannel interfaces may also lead to more robust designs for nanofluidic systems~\cite{schoch2008}, e.g. for DNA or protein sequencing or molecular sorting, in this case to inhibit the formation of shocks, which interfere with external control of dynamics within the nanochannel. 

Another interesting direction would be to relax the assumption of fixed surface charge, and allow for capacitive charging~\cite{biesheuvel2010}, Faradaic reactions~\cite{newman1975,biesheuvel2011}, or induced-charge electro-osmotic  flows~\cite{cocis2010} in microfluidic devices or porous electrodes. Leinweber et al.~\cite{leinweber2006} have observed that metal micropost arrays in thin (1 micron) channels can produce strong concentration polarization and continuous desalination. The effect is driven by surface conduction on ideally polarizable metal cylinders~\cite{chu2006}. It is likely that desalination shock phenomena, due to surface conduction on the microchannel walls, also play a role in shaping the salt concentration profile in these devices. 

In the case of porous electrodes, our volume-averaged equations for porous media can be applied to  capture effects of surface conduction, but they must be augmented by a charge-voltage relation for the double layer, e.g. using the Gouy-Chapman-Stern model of capacitive charging~\cite{bazant2004,biesheuvel2010} or the Frumkin-Butler-Volmer-Stern model of Faradaic reactions~\cite{biesheuvel2009,biesheuvel2011}.  Porous electrodes are widely used in electrochemical energy storage  devices (batteries, supercapacitors, fuel cells, etc.)~\cite{newman1975,newman2004}, but we are not aware of any prior work considering surface conduction. Designing the porous microstructure to exploit the nonlinear effects of surface conduction could provide a new means to enhance the power density of portable power sources.

%\appendix
%\section{}

\appendix
\section{Porous media with nonuniform properties}
\label{appendixA}

 In this appendix we present a more general form of Eqs.~\ref{eq:ceq},~\ref{eq:jeq}, and~\ref{eq:jb} applicable to porous media with nonuniform properties such as porosity, diffusivity, and area density. We here allow for variable diffusivities, independent of mobility (no Einstein relation). Variable diffusivity can be due to variable geometrical properties of the microstructure or due to nonlinear flow dispersion effects which enhances the effective diffusivity in the flow direction~\cite{probstein1994}.  The effect of Taylor dispersion due to electro-osmotic flow has been analyzed for thin capillaries~\cite{ghosal2003} and flat microchannels~\cite{storey2005}, and accurate volume-averaged equations are available for these situations. Yaroschuk and Zholkovskiy~\cite{yaroshchuk} have recently predicted that this effect  can also produce sharp fronts in the salt concentration in a microchannel, near a nanochannel junction, although mainly in thicker microchannels (around $100 \mu$m)~\cite{yaroshchuk}. While the following model would accommodate such effects, we here briefly note that a simple scaling argument suggests that Taylor dispersion can be neglected in very thin ($h_p<\sim\mu$m) microstructures due to their relatively low P\'eclet number, $\mbox{Pe} = uh_p/D$~\cite{dydek}. 

 To derive the model we start with the general form of Eq.~\ref{eq:binnp}
\begin{equation*}
\frac{\partial \epsilon_pc_\pm}{\partial t} + \nabla\cdot(\epsilon_p\ub c_\pm) = \nabla\cdot\left[ \epsilon_pD_\pm\nabla c_\pm \pm \epsilon_pz_\pm\nu_\pm c_\pm kT\nabla \tilde{\phi}\right],
\label{eq:binnpgen} 
\end{equation*}
where we remind that $\epsilon_p$ is the porosity of the porous medium. Higher porosity indicates higher effective volume to accommodate the transport and thus all fluxes scale proportionally with porosity. In this case the conservation laws need to be weighted by the local porosity factors. For example, the continuity equation would be $\nabla\cdot(\epsilon_p\ub)=0$  instead of $\nabla\cdot\ub=0$, etc. Rewriting  Eq. \ref{eq:binnpgen} in terms of $c_b$, defined by Eq.~\ref{eq:cb}, and using net neutrality (see Eq.~\ref{eq:neutrality}) results in 
\begin{equation}
\frac{\partial \epsilon_pc_b}{\partial t} + \nabla\cdot(\epsilon_p\ub c_b) = \nabla\cdot\left[ \epsilon_p\overline{D}(\nabla c_b - \frac{\bar{z}}{e}\rho_s\nabla\tilde{\phi}) +{\bf f}_s\right], \label{eq:ceqgen} 
\end{equation}
\begin{equation}
0 = \nabla\cdot(\epsilon_p\jb)   \label{eq:jeqgen},
\end{equation}
where,
\begin{equation}
\overline{D}=\frac{z_-\nu_-D_++z_+\nu_+D_-}{z_-\nu_-+z_+\nu_+}, \nonumber 
\end{equation}
\begin{equation}
\overline{z}=\frac{2z_+z_-\nu_+\nu_-kT}{z_-D_+\nu_-+z_+D_-\nu_+}. 
\end{equation}
The ${\bf f}_s$ flux appears as a consequence of nonuniform surface charge, $\rho_s$ and is equal to 
\begin{equation}
{\bf f}_s=\frac{\epsilon_p}{e}\frac{2z_-\nu_-}{z_+\nu_+ + z_-\nu_-}\left(\rho_s \ub-D_+\nabla \rho_s\right).
\end{equation}

To close the system of Eqs~\ref{eq:ceqgen} and \ref{eq:jeqgen} we introduce the relation between current and potential gradient, by updating Eq. 10
\begin{equation}
\frac{e}{kT}(\jb + \rho_s \ub - D_+\nabla \rho_s) = - \beta  \nabla \kappa_b - \left[\kappa_b + \kappa_s/h_p\right] \nabla \tilde{\phi},   \label{eq:jbgen}
\end{equation}
which only has a slight modification relative to Eq.~\ref{eq:jb} due to nonuniformity of $\rho_s$ with $\beta$, $\kappa_b$, and $\kappa_s$ defined the same as in the main text.

\section{Desalination shock profile in general microstructures}
\label{appendixB}
Here we analyze shock structure in a microtubal structure whose area $a(x)$ and perimeter $p(x)$ vary independently with position. Due to the mathematical equivalence of microtubes and porous structures in our model, the same analysis also holds for porous medium with variable porosity $\epsilon_p(x)$ and surface area density $a_p(x)$, which respectively play analogous roles as $a$ and $p$ here. We start with the nondimensional version of Eq.~\ref{eq:general}, where we 
use $a_0$ and $p_0$, respectively the channel cross-sectional area and perimeter evaluated at $x_0$, to nondimensionalize $a$ and $p$. 

Using the other dimensionless variables from the main text, we arrive at the following dimensionless equation describing evolution of bulk conductivity in a channel with gradually varying $a(x)$ and $p(x)$:
\begin{equation}
 \frac{\partial}{\partial \tilde{t}}(\tilde{a}\tilde{\kappa})+\frac{\partial}{\partial \tilde{x}}\left(\tilde{u}\tilde{\kappa}+\frac{\tilde{p}\Du}{\tilde{a}\tilde{\kappa}+\tilde{p}\Du}\right)=\frac{\partial}{\partial \tilde{x}}\left[\tilde{a}\frac{\partial \tilde{\kappa}}{\partial \tilde{x}}\right],
 \label{eq:dimensionlessPA}
\end{equation}
where $\tilde{\kappa}=\kappa_b/\kappa_{b\infty}$, $\tilde{x}=(x/\overline{D})(z_-\nu_-ej_0/\kappa_b\infty)$, $\tilde{t}=(t/\overline{D})(z_-\nu_-ej_0/\kappa_b\infty)^2$, and $\tilde{u}=u_0\kappa_{b\infty}/z_-\nu_-ej_0$. To include a more general case with gradual variation of surface conductivity, we define $\Du=p_0\kappa_{s0}/a_0\kappa_{b\infty}$; in this case $\tilde{p}$ represents variation of both surface charge and perimeter and is defined as $\tilde{p}=p\kappa_s/p_0\kappa_{s0}$.
\par \ \par
We assume that the changes in $\tilde{a}$ and $\tilde{p}$  are slow enough, so that their variation over the shock can be neglected. We use $\tilde{\kappa}_1$ and $\tilde{\kappa}_2$ to denote respectively the left and right conductivities out side the shock, but close enough so that the cross-section is the same as that at the shock. Therefore $\tilde{\kappa}_1$ and $\tilde{\kappa}_2$ may vary as the shock sweeps through the channel, which later will be obtained from quasi-steady solutions.
\par \ \par
If the shock structure moves with local velocity $\tilde{v}$, following the transformation $\tilde{y}=\tilde{x}-\tilde{v}\tilde{t}$ we obtain the following ODE governing structure of the shock.
\begin{equation}
 \frac{d}{d \tilde{y}}\left(\tilde{\kappa}(\tilde{u}-\tilde{a}\tilde{v})+\frac{\tilde{p}\Du}{\tilde{a}\tilde{\kappa}+\tilde{p}\Du}\right)=\frac{d}{d \tilde{y}}\left[\tilde{a}\frac{d \tilde{\kappa}}{d \tilde{y}}\right].
 \label{eq:shock}
\end{equation}
Integration yields
\begin{equation}
\tilde{\kappa}(\tilde{u}-\tilde{a}\tilde{v})+\frac{\tilde{p}\Du}{\tilde{a}\tilde{\kappa}+\tilde{p}\Du}=\tilde{a}\frac{d \tilde{\kappa}}{d \tilde{y}}+C.
 \label{eq:shock2}
\end{equation}
We use $\tilde{\kappa}_1$ and $\tilde{\kappa}_2$ as the boundary condition at infinity. Evaluating Eq. \ref{eq:shock2} at $\pm\infty$ and ignoring the diffusion term yields the values of $C$ and $\tilde{v}$:
\begin{equation}
(\tilde{u}-\tilde{a}\tilde{V})=\frac{\tilde{a}\tilde{p}\Du}{(\tilde{a}\tilde{\kappa}_2+\tilde{p}\Du)(\tilde{a}\tilde{\kappa}_1+\tilde{p}\Du)},
\label{eq:V}
\end{equation}
\begin{equation}
C=\frac{\tilde{p}\Du(\tilde{a}\tilde{\kappa}_2+\tilde{a}\tilde{\kappa}_1+\tilde{p}\Du)}{(\tilde{a}\tilde{\kappa}_2+\tilde{p}\Du)(\tilde{a}\tilde{\kappa}_1+\tilde{p}\Du)}.
 \label{eq:C}
\end{equation}
Substituting into Eq. \ref{eq:shock2} yields:
\begin{equation}
\frac{\tilde{a}\tilde{p}\Du}{(\tilde{a}\tilde{\kappa}_2+\tilde{p}\Du)(\tilde{a}\tilde{\kappa}_1+\tilde{p}\Du)}\frac{(\tilde{\kappa}-\tilde{\kappa}_2)(\tilde{\kappa}-\tilde{\kappa}_1)}{\tilde{a}\tilde{\kappa}+\tilde{p}\Du}=\frac{d\tilde{\kappa}}{d\tilde{y}}.
 \label{eq:shock3}
\end{equation}
Rearranging terms yields:
\begin{equation}
\frac{\tilde{a}\tilde{p}\Du d\tilde{y}}{(\tilde{a}\tilde{\kappa}_2+\tilde{p}\Du)(\tilde{a}\tilde{\kappa}_1+\tilde{p}\Du)}=-\frac{\tilde{a}\tilde{\kappa}_1+\tilde{p}\Du}{\tilde{\kappa}_1-\tilde{\kappa}_2}\frac{d\tilde{\kappa}}{\tilde{\kappa}_1-\tilde{\kappa}} - \frac{\tilde{a}\tilde{\kappa}_2+\tilde{p}\Du}{\tilde{\kappa}_1-\tilde{\kappa}_2}\frac{d\tilde{\kappa}}{\tilde{\kappa}-\tilde{\kappa}_2}.
\label{eq:shock4}
\end{equation}
Integration results in
\begin{equation}
\frac{\tilde{a}\tilde{p}\Du (\tilde{y}-\tilde{y}_0)}{(\tilde{a}\tilde{\kappa}_2+\tilde{p}\Du)(\tilde{a}\tilde{\kappa}_1+\tilde{p}\Du)}=\frac{\tilde{a}\tilde{\kappa}_1+\tilde{p}\Du}{\tilde{\kappa}_1-\tilde{\kappa}_2}\text{ln}(\tilde{\kappa}_1-\tilde{\kappa}) - \frac{\tilde{a}\tilde{\kappa}_2+\tilde{p}\Du}{\tilde{\kappa}_1-\tilde{\kappa}_2}\text{ln}(\tilde{\kappa}-\tilde{\kappa}_2).
 \label{eq:shock5}
\end{equation}
Now we need to substitute values of $\tilde{\kappa}_1$ and  $\tilde{\kappa}_2$ in terms $\tilde{\kappa}_d$ and local $\tilde{p}$ and $\tilde{a}$.  $\tilde{\kappa}_2$ satisfies the steady state condition for Eq. \ref{eq:dimensionlessPA} in the depletion region. Since we are far from the shock the diffusive flux can be neglected in this region; hence the net convective flux should be constant in order to satisfy the steady state condition. Therefore,
\begin{equation}
\tilde{\kappa}_2+\frac{\tilde{p}\Du}{\tilde{a}\tilde{\kappa}_2+\tilde{p}\Du}=\tilde{\kappa}_d+\frac{\Du}{\tilde{\kappa}_d+\Du}.
\end{equation}
Note that $\tilde{p}$ and $\tilde{a}$ are one at $\tilde{x}=\tilde{x}_0$. Considering the fact that $\tilde{\kappa}_2\sim \tilde{\kappa}_d\sim O(\Du)\ll 1$, we can simplify this expression and arrive at
\begin{equation}
\tilde{\kappa}_2=\frac{\tilde{\kappa}_d\tilde{p}}{\tilde{a}}+O(\Du^2).
\end{equation}
Similarly, one can show that
\begin{equation}
\tilde{\kappa}_1=1+O(\Du).
\end{equation}
Substituting these expressions for $\tilde{\kappa}_1$ and $\tilde{\kappa}_2$ into Eq. \ref{eq:shock5} results in
\begin{equation}
\left(\frac{\Du}{\tilde{\kappa}_d+\Du}\right)\frac{\tilde{y}-\tilde{y}_s}{\tilde{a}}=\text{ln}(1-\tilde{\kappa}) - (\tilde{\kappa}_d+\Du)(\tilde{p}/\tilde{a})\text{ln}\left[\tilde{\kappa}-\tilde{\kappa}_d(\tilde{p}/\tilde{a})\right] ,
\label{eq:shock6}
\end{equation}
which is a direct relation between the bulk conductivity and axial coordinate across a shock. Having $\tilde{x}_s=\tilde{y}_0+\tilde{v}\tilde{t}$ this equation can be transformed to Eq.~\ref{eq:shockprofile}. 

Figure~\ref{fig4}C shows the shock profiles obtained from Eq. \ref{eq:shock6}. One can see that different regions of the shock scale differently as parameters $\tilde{a}$ and $\tilde{p}$ vary. While the high-concentration region of the shock scales with local $\tilde{a}$, the low-concentration region is dependent on both parameters $\tilde{a}$ and $\tilde{p}$. This also makes sense from the form of Eq. \ref{eq:shock6} since the high- and low-concentration regions can be approximated respectively by the first and second term in the right hand side of the Eq. \ref{eq:shock6}. A plot of the shock profile together with these two asymptotic profiles are shown in Fig.~\ref{fig4}.

In practical scenarios the conductivity-drop across the shock is orders of magnitude ($O(\Du)\ll 1$). Under such conditions most of the drop, from $\tilde{\kappa}=1$ to $\Du\ll\tilde{\kappa}\ll 1$, can be approximated by only the first term on the right-hand-side of Eq. \ref{eq:shock6}. Therefore, as a rule of thumb, one can say that the shock thickness approximately scales with local area. Note that this simple criterion  assumes that variations in $\tilde{p}/\tilde{a}$ are finite and bounded with an upper bound much smaller than $1/\Du$. 

\begin{acknowledgments}
This work was supported in part by the MIT Energy Initiative.  The authors thank S. S. Dukhin and N. A. Mishchuk for important references.
\end{acknowledgments}

%\bibliographystyle{apsrev4-1}
\bibliography{bibitems}

\begin{thebibliography}{86}
\bibitem[{mov()}]{movie}
\bibinfo{note}{A movie is included in the supporting online material. The ions
  diffuse between hard walls according to Brownian dynamics and drift in the
  mean electric field. The latter is determined self consistently by solving
  Poisson's equation with a smeared-out (locally volume averaged) ionic charge
  density and a constant charge density on the walls.}
\end{thebibliography}

\end{document}